# A New Image Steganography Based On First Component Alteration Technique

Amanpreet Kaur[1], Renu Dhir[2], and Geeta Sikka[3]

[1,2,3] Department of Computer Science and Engineering. National Institute of Technology, Jalandhar, India

*Abstract*—In this paper, A new image steganography scheme is proposed which is a kind of spatial domain technique. In order to hide secret data in cover-image, the first component alteration technique is used. Techniques used so far focuses only on the two or four bits of a pixel in a image (at the most five bits at the edge of an image) which results in less peak to signal noise ratio and high root mean square error. In this technique, 8 bits of blue components of pixels are replaced with secret data bits. Proposed scheme can embed more data than previous schemes and shows better image quality. To prove this scheme, several experiments are performed, and are compared the experimental results with the related previous works.

## I. INTRODUCTION

Due to rapid development in both computer technologies and Internet, the security of information is regarded as one of the most important factors of Information Technology and communication. Accordingly, we need to take measures which protect the secret information.

Generally, secret information may be hidden in one of two ways, such as cryptography and steganography. The methods of cryptography makes the data unintelligible to outsiders by various transformations, whereas the methods of steganography conceal the existence of messages. The word steganography is derived from Greek words meaning "covered writing" and as such it centers on the concept of hiding a message. As defined by Cachin [3], steganography is the art and science of communicating in such a way that the presence of message is detected. Steganography is very old method used. Around 440 B.C., Histiaeus shaved the head of his most trusted slave and tattooed it with a message which disappeared after the hair had regrown. The purpose of this message was to instigate a revolt against the Persians. Another slave could be used to send a reply. During the American Revolution, invisible ink which would glow over a flame was used by both the British and Americans to communicate secretly.

Steganography was also used in both World Wars. German spies hid text by using invisible ink to print small dots above or below letters and by changing the heights of letter-strokes in cover texts.

Among the methods of steganography, the most common thing is to use images for steganography. This is called image steganography. In this image hiding method, the pixels of images are changed in order to hide the secret data so as not to be visible to users, and the changes applied in the image are not tangible. The image used to camouflage the secret data is called the cover-image while the cover-image with the secret data embedded in it is called the stego-image.

Image steganographic techniques can be divided into two groups [7]: the Spatial Domain technique group, and the Transform Domain technique group. The Spatial domain technique embeds information in the intensity of the pixels directly, while the Transform domain technique embeds information in frequency domain of previously transformed image. Our proposed scheme is a kind of the spatial domain techniques.

## II. RELATED WORKS

### A. Least Significant Bit Hiding (LSB) Scheme

This method is probably the easiest way of hiding information in an image. In the LSB technique, the LSB of the pixels is replaced by the message to be sent. The message bits are permuted
Before embedding, which has the effect of distributing the bits evenly, thus on average only half of the LSB's will be modified.

### B. Pixel-Value Differencing (PVD) scheme

The alteration of edge areas in the human visual system cannot be distinguished well, but the alteration of smooth areas can be distinguished well. That is, an edge area can hide more secret data than a smooth area. With this concept, Wu and Tsai proposed a novel steganography technique using the pixel-value differencing (PVD) method to distinguish edge and smooth areas. The PVD technique can embed more data in the edge area which guarantees high imperceptibility.

### C. Lie-Chang's scheme

The Steganographic technique has to possess two important properties. These are good imperceptibility and sufficient data capacity. A scheme which satisfied both properties was proposed by Lie-Chang [5]. The scheme is an Adaptive LSB technique using Human Visual System (HVS). HVS has the following characteristics: Just Noticeable Difference (JND), Contrast Sensitivity Function (CSF), Masking and Spectral

Sensitivity. The characteristic of HVS used by Lie-Chang is JND (also known as the visual increment threshold or the luminance difference threshold). In this scheme, JND is defined as the amount of light $\Delta I$ necessary to add to a visual field of intensity $I$ such that it can be distinguished from the background. In HVS, the curve for $\Delta I$ versus $I$ can be analytically and mathematically modeled.

The JND technique is simple and has a higher embedding capacity than other schemes. Also, this technique has high embedding capacity about overall bright images and has high distortion of a cover image when the embedding capacity is increased, but does not concern overall dark images.

### D. MSB3 edge-detection

Generally, the human eye is highly sensitive to overall pictures of the field of view, while having low senstivity to fine details. Such characteristic of HVS is called the CSF. One of several computational models which explain the CSF is proposed by Mannos-Sakrison [1]. According to Mannos-Sakrison scheme, if additional data is embedded in the pixels of high spatial frequency, one is able to satisfy both the increment of hiding capacity and good imperceptibility.

In order to judge whether any pixel has the high spatial frequency or the low spatial frequency in a digital image, the edge detection algorithm is generally used. The GAP algorithm is one of the edge detection algorithms. For the input value of the GAP algorithm, we use the technique of the using three bits from the MSB. Due to this, the pixels that are selected from an embedding phase must be equal to pixels that are selected from an extracting phase. Three bits are embedded in a pixel if the pixel-value is smaller than the first threshold value (intensity 88) and is judged with the edge-region. MSB3 Edge-Detection is summarized through the following steps:

**Step 1:** Execute MSB3 Edge-Detection at a cover image, in order to sort out edge- regions in the
cover image.

**Step 2:** If any pixel value is smaller than the first threshold value and exists on the edge-
region, embed three bits of secret data in the pixel.

### E. Image Steganography Based on $2^k$ Correction and Edge- Detection Scheme

In this method, author used the just noticeable difference (JND) technique and method of contrast sensitivity function (CSF). This is an MSB3 edge-detection which uses part information of each pixel-value. In order to have better imperceptibility, a mathematical method which is the $2^k$ correction is used.

If one supposes the secret data is hidden at a pixel of cover image, some differences occurred between cover-pixel and stego-pixel. Because of these differences, the cover image is distorted and the quality of cover image is dropped. $2^k$ correction corrects each pixel-value as $2^k$. That is, supposing that $k$-bits are embedded in a pixel value, the method *adds* or *subtracts* $2^k$ to each pixel-value, and finally the corrected pixel value becomes closer to the original-pixel. Hence, the secret data in the stego-pixel is not changed.

This scheme can embed more data than previous schemes, and shows better imperceptibility. The method is an edge-detection which only uses 3-bits from MSB of each pixel-value. In this method, data embedment depends on the value received from each pixel value whether it is on the edge or on other part of an image. If it is on the edge it embed data in the cover image based on the value of k, value of k is decided on the pixel position whether it is on the edge or not. This method modifies the stego pixel value near to the cover pixel using 2k correction mathematical formula.

### III. PROPOSED IMAGE STEGANOGRAPHY SCHEME

In the proposed scheme, a new image steganography scheme based on first componenet alteration technique is introduced. In a computer, images are represented as arrays of values. These values represent the intensities of the three colors R (Red), G (Green) and B (Blue), where a value for each of three colors describes a pixel. Each pixel is combination of three components(R, G, and B).

In this scheme, the bits of first component (blue component) of pixels of image have been replaced with data bits, which are applied only when valid key is used. Blue channel is selected because a research was conducted by Hecht, which reveals that the visual perception of intensely blue objects is less distinct that the perception of objects of red and green.

For example, suppose one can hide a message in three pixels of an image (24-bit colors). Suppose the original 3 pixels are:

($00100111$  11101001  11001000) (00100111  11001000 11101001)
(11001000 00100111 11101001)

A steganographic program could hide the letter "A" which has a position 65 into ASCII character set and have a binary representation "01000001", by altering the blue channel bits of pixels.

($01000001$  11101001  11001000) (00100111  11001000 11101000)
(11001000 00100111 11101001)

### A. Embedding phase

The embedding process is as follows.
Inputs: Image file and the text file
Output: Text embedded image

Procedure:
Step 1: Extract all the pixels in the given image and store it in the array called Pixel-Array.
Step 2: Extract all the characters in the given text file and store it in the array called Character- Array.
Step 3: Extract all the characters from the Stego key and store it in the array called Key- Array.
Step 4: Choose first pixel and pick characters from Key-Array and place it in first component of pixel. If there are more characters in Key- Array, then place rest in the first component of next pixels, otherwise follow Step (e).
Step 5: Place some terminating symbol to indicate end of the key. '0' has been used as a terminating symbol in this algorithm.

Step 6: Place characters of Character- Array in each first component (blue channel) of next pixels by replacing it.
Step 7: Repeat step 6 till all the characters has been embedded.
Step 8: Again place some terminating symbol to indicate end of data.
Step 9: Obtained image will hide all the characters that we input.

*B. Extraction phase*
The extraction process is as follows.
Inputs: Embedded image file
Output: Secret text message
Procedure:
Step 1: Consider three arrays. Let they be Character-Array, Key-Array and Pixel-Array.
Step 2: Extract all the pixels in the given image and store it in the array called Pixel-Array.
Step 3: Now, start scanning pixels from first pixel and extract key characters from first (blue) component of the pixels and place it in Key-Array. Follow Step 3 till we get terminating symbol, otherwise follow step 4.
Step 4: If this extracted key matches with the key entered by the receiver, then follow Step 5, otherwise terminate the program by displaying message "Key is not matching".
Step 5: If the key is valid, then again start scanning next pixels and extract secret message characters from first (blue) component of next pixels and place it in Character Array. Follow Step 5 till we get terminating symbol, otherwise follow step 6.
Step 6: Extract secret message from Character-Array.

## IV. EXPERIMENTAL RESULTS

To evaluate the performance of the proposed scheme, the Image Steganography is firstly applied to Lena's image as a test image. Different results have been observed with RGB components by changing first component to embed data in it and to measure image quality of the proposed scheme, we used Peak Signal-to-Noise Ratio (PSNR) and the MSE (Mean Square Error) for an Encrypted Image. The results are then compared with various Encryption Method as shown in the table. The PSNR computes the peak signal-to-noise ratio, in decibels, between two images. This ratio is often used as a quality measurement between the original and a compressed image. The higher the PSNR, the better the quality of the compressed or reconstructed image. The MSE (Mean Square Error) represents the cumulative squared error between the compressed and the original image, the lower the value of MSE, the lower the error.
To compute the PSNR, the block first calculates the mean-squared error using the following equation:

$$MSE = \frac{1}{M*N} \sum_{x=1}^{M} \sum_{y=1}^{N} [x(m,n) - y(m,n)]^2$$

Where x (m, n) and y (m, n) are the two images of size m*n. In this case x is the original image and y is the encrypted image.

$$PSNR = 20 \log_{10} \left[ \frac{MAXPIX}{RMSE} \right]$$

Where MAXPIX is the maximum pixel value and RMSE is the Root Mean Square Error of the image (it quantifies the average sum of distortion in each pixel of the encrypted image i.e. average change in pixel caused by encryption algorithm)

$$RMSE = \sqrt{MSE}$$

The results are then compared with various steganography methods as shown in the following table.

| Lena image | LSB3 | PVD | Lie Chang's | Jae Gil Yu | First Component alteration technique |
|---|---|---|---|---|---|
| PSNR | 37.92 | 41.48 | 37.53 | 38.98 | 46.11 |

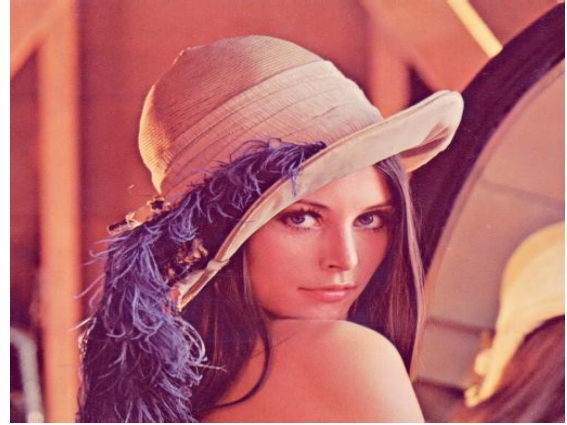

Figure 2(a) Original Lena image (512 X 512)

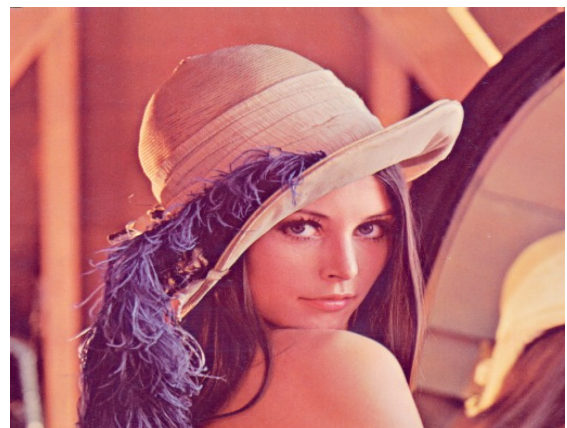

Figure 2(b) Stego Lena Image (512 X 512)

## V. CONCLUSIONS

In this paper, a new image steganography scheme which is a kind of spatial domain technique is used. In order to hide secret data in cover-image, the first component alteration technique is used. Techniques used so far focuses only on the two or four bits of a pixel in a image ,(at most five bits at the edge of an image.) which results less peak to signal noise ratio and high root mean square error i.e. less than 45 PSNR value. Proposed work is concentrated on 8 bits of a pixel (8 bits of blue component of a randomly selected pixel in a 24 bit image), resulting better image quality. Proposed technique has also used contrast sensitivity function (CSF) and just noticeable difference (JND) Model. Proposed scheme can embed more data than previous schemes [7, 5, 10], and shows better imperceptibility. To prove this scheme, several experiments are performed, and the experimental results are compared with the related previous works. Consequently, the experimental results proved that the proposed scheme is superior to the related previous works.

The future work is to extend proposed technique for videos and to modify given scheme to improve image quality by increasing PSNR value and lowering MSE value.